\documentclass[twocolumn, prd, showpacs, preprintnumbers]{revtex4}
\usepackage{CJK}

\begin{document}

\title{Comment on astrophysical consequences of a neutrinophilic 2HDM}

\author{Shun Zhou}

\email{zhoush@mppmu.mpg.de}

\affiliation{Max-Planck-Institut f\"ur Physik
(Werner-Heisenberg-Institut), F\"ohringer Ring 6, 80805 M\"unchen,
Germany}

\date{\today}
\preprint{MPP-2011-70}

\begin{abstract}
Several authors have pointed out that the scalar-mediated
interaction of neutrinos in a neutriophilic two-Higgs-doublet model
($\nu$2HDM) can be as strong as electromagnetic interaction
\cite{Sher:2011mx,Wang:2006jy,Gabriel:2006ns}. We show that the
coupling constants of neutrino-scalar interaction are actually
restricted to be $y_i \lesssim 1.5\times 10^{-3}$ by supernova
neutrino observation, and further constrained to be $y_i \lesssim
2.3\times 10^{-4}$ by precision measurements of acoustic peaks of
the cosmic microwave background. Based on the energy-loss argument
for supernova cores, we derive a slightly more restrictive bound
$y_i \lesssim 3.5 \times 10^{-5}$. Therefore, the $\nu$2HDM has lost
its spirit of explaining tiny Dirac neutrino masses while keeping
neutrino Yukawa couplings of order one.
\end{abstract}
%

\maketitle

The $\nu$2HDM extends the standard model with three right-handed
neutrinos and introduces exclusively for neutrinos an extra Higgs
doublet $\varphi$, which acquires a small vacuum expectation value
(vev) $v_\varphi = 0.1~{\rm eV}$. In this model, neutrinos are Dirac
particles, and their mass matrix is given by $M_\nu = Y_\nu
v_\varphi$ with $Y_\nu$ being neutrino Yukawa coupling matrix. A
salient feature of the $\nu$2HDM is that $Y_\nu$ can be of order one
even for sub-eV neutrino masses
\cite{Sher:2011mx,Wang:2006jy,Gabriel:2006ns}.
In the flavor basis where the charged-lepton Yukawa coupling matrix
is diagonal, one can further take $Y_\nu$ to be Hermitian by
rotating right-handed neutrino fields in the flavor space. Hence the
Yukawa interaction of neutrinos can be written as
\begin{equation}
-{\cal L}_{\rm Y} = \sum^\tau_{\alpha,\beta=e}
\left(Y_\nu\right)_{\alpha \beta} \overline{\nu_\alpha} \nu_\beta
\eta = \sum^3_{i=1} y_i \overline{\nu_i} \nu_i \eta \; ,
\end{equation}
where $V^\dagger Y_\nu V = {\rm Diag}\{y_1, y_2, y_3\}$ with $V$
being the neutrino mixing matrix, which relates neutrino mass
eigenstates $\nu_i$ to flavor eigenstates $\nu_\alpha$, and $m_i =
y_i v_\varphi$ (for $i=1, 2, 3$) are neutrino masses. Here $\eta$ is
a scalar boson arising from the neutral component of $\varphi$, and
its mass is naturally around the vev of $\varphi$, i.e., $m_\eta
\approx v_\varphi = 0.1~{\rm eV}$. Some cosmological and
astrophysical consequences of the neutrino interaction in Eq. (1)
with $y_i \sim {\cal O}(1)$ have been discussed in Refs.
\cite{Sher:2011mx,Wang:2006jy,Gabriel:2006ns}, however, the
restrictive bounds on neutrino Yukawa couplings are unfortunately
missed.

In fact, stringent bounds on the neutrino-Majoron interaction have
been obtained in the literature by assuming a pseudoscalar coupling
$i y_i \overline{\nu_i}\gamma_5 \nu_i \chi$ with $\chi$ being a
pseudoscalar boson \cite{Kolb:1987qy,Hannestad:2005ex}. One can show
that those bounds apply as well to the scalar case in the
relativistic limit, where small neutrino masses can be neglected.
However, it should be noticed that the lepton-number-violating
processes are forbidden in the $\nu$2HDM.

The observation of neutrinos from Supernova 1987A requires that the
mean free path of electron antineutrinos in the presence of cosmic
background particles should be larger than the supernova distance,
i.e., $\lambda_{\bar\nu_e}^{-1} D \lesssim 1$ with $D = 51.4~{\rm
kpc}$, in order to avoid significant reduction of neutrino flux
\cite{Kolb:1987qy}. The relevant processes are $\bar\nu_e + \eta \to
\bar\nu_e + \eta$, $\bar\nu_e + \nu_e \to \eta + \eta$ and
$\bar\nu_e + \nu_\alpha \to \nu_e + \bar\nu_\alpha$ for $\alpha = e,
\mu, \tau$. After removing the lepton-number-violating contributions
from the Majoron model \cite{Kolb:1987qy}, one can obtain a
restrictive bound on neutrino Yukawa couplings
\begin{equation}
y_i \lesssim 1.5\times 10^{-3} \; .
\end{equation}
If $\eta$ bosons decay rapidly into neutrinos and are absent in the
cosmic background, the bound will be weaker but on the same order of
magnitude. Given $v_\varphi = 0.1~{\rm eV}$, neutrino Yukawa
couplings are determined by neutrino masses $y_i = m_i/v_\varphi$,
thus heavier neutrino mass eigenstates interact more strongly with the
scalar boson. In the case of normal mass hierarchy, the bound in
Eq. (2) may be slightly relaxed to $y_i \lesssim 10^{-2}$, because
electron antineutrino possesses a small fraction of the heaviest mass
eigenstate\cite{Sher}. For the inverted mass hierarchy or nearly
degenerate neutrino mass spectrum, however, the bound in Eq. (2) is
still applicable.

As indicated by precision measurements of the acoustic peaks of the
cosmic microwave background, neutrinos should be freely streaming
around the time of photon decoupling $T_{\gamma,{\rm dec}} =
0.256~{\rm eV}$ in order to avoid the acoustic oscillations of the
neutrino-scalar fluid \cite{Hannestad:2005ex}. At this moment, the
neutrino temperature is $T_{\nu,{\rm dec}} = (4/11)^{1/3} T_{\gamma,
{\rm dec}} = 0.183~{\rm eV}$. For the relevant two-body scattering
processes $\nu_i + \eta \to \nu_i + \eta$, $\nu_i + \bar\nu_i \to
\eta + \eta$ and $\nu_i + \bar\nu_i \to \nu_j + \bar\nu_j$ via
$\eta$-exchange, we can simply estimate the scattering rate as
$\Gamma_\nu \approx y^4_i T_{\nu, {\rm dec}}$ up to some numerical
factors. The free-streaming argument requires $\Gamma_\nu$ to be
smaller than the cosmic expansion rate $H_{\gamma, \rm dec} =
100~{\rm km}~{\rm s}^{-1}~{\rm Mpc}^{-1} (\Omega_{\rm M} h^2)^{1/2}
(z_{\rm dec} + 1)^{3/2}$ with $\Omega_{\rm M} h^2 = 0.134$ being the
cosmic matter density and $z_{\rm dec} = 1088$ the redshift at
photon decoupling. Hence one can derive a more restrictive bound
$y_i \lesssim 1.1\times 10^{-7}$ \cite{Hannestad:2005ex}. Taking
account of the existing bound on Yukawa couplings in Eq. (2), one
should increase $v_\varphi$ by three orders of magnitude to guarantee
sub-eV neutrino masses. Therefore, the mass of $\eta$ is expected to
be $m_\eta \approx v_\varphi = 100~{\rm eV}$, which is much larger
than neutrino temperature at the time of photon decoupling. As a
consequence, $\eta$ bosons have already decayed into neutrinos, and
the relevant process $\nu_i + \bar\nu_i \to \nu_j + \bar\nu_j$ is
mediated by a virtual $\eta$ boson \cite{Sher}. The scattering rate
is modified to be $\Gamma_\nu \approx y^4_i T^5_{\nu, {\rm
dec}}/m^4_\eta \approx y^4_i T^5_{\nu, {\rm dec}}/v^4_\varphi =
y^8_i T^5_{\nu, {\rm dec}}/m^4_i$, thus the true bound from the
free-streaming argument is
\begin{equation}
y_i \lesssim 2.3 \times 10^{-4} \; ,
\end{equation}
for $m_i \sim 0.1~{\rm eV}$. Since the scattering rate is
proportional to $y^8_i$, the neglected numerical factors are indeed
unimportant for the bound in Eq. (3).

An important point is that $\eta$ is massive enough to decay into
neutrino-antineutrino pairs $\eta \to \nu_i + \bar\nu_i$. The
lifetime in its rest frame is $\tau_\eta = (3y^2_i
m_\eta/16\pi)^{-1} \approx 1.1\times 10^{-9}~{\rm s}$, where $y_i =
10^{-4}$ and $m_\eta \approx v_\varphi = 1~{\rm keV}$ have been
taken. Although $\eta$ bosons can be copiously produced in the
supernova core, one may expect that they will decay soon after
production and thus cannot cause excessive energy losses. However,
since the temperature in the cooling phase is sufficiently high $T =
30~{\rm MeV}$, the lifetime of thermal $\eta$ bosons should be
lengthened by a Lorentz factor $E/m_\eta \approx 10^5$.
Consequently, the relativistic $\eta$ bosons before decaying may
have traveled a distance $l_\eta \approx 3.3\times 10^6~{\rm cm}$,
which is larger than the core radius $R = 10~{\rm km}$. But $\eta$
bosons cannot freely propagate in the neutrino background, their
mean free path can be estimated as $\lambda_\eta = (y^4_i T)^{-1}
\approx 6.6\times 10^3~{\rm cm}$ that is comparable to the mean free
path of neutrinos in the case of standard neutral-current
interaction. Thus $\eta$ bosons behave like a new species of
neutrinos and accelerate the energy transfer, which leads to the
reduction of the cooling time or the duration of supernova neutrino
burst. There are two possibilities to avoid the contradiction with
the neutrino observation of Supernova 1987A \cite{Raffelt:1990yz}:
(i) to increase the coupling ($y_i > 10^{-4}$) such that
$\lambda_\eta$ becomes much smaller and the energy transfer by
$\eta$ bosons is negligible; (ii) to decrease the coupling ($y_i <
10^{-4}$) such that $\eta$ bosons have never been trapped and
thermalized in the core. The first possibility is already excluded
by the bound in Eq. (3), while the second one is subject to the
constraint from standard energy-loss arguments.

The production of $\eta$ bosons will be efficient via the
bremsstrahlung process $\nu_i + N \to \nu_i + N + \eta$ because of
the high nucleon density. The emission rate should be proportional
to the thermal average of $\nu N$ scattering rate $\sigma n_\nu
n_{\rm B}$ where $\sigma$ is the $\nu N$ collision cross section,
$n_\nu$ and $n_{\rm B}$ are respectively the neutrino and baryon
number densities. The energy of emitted $\eta$ bosons is of order
$T$. Put all together, we can get the volume emission rate
\begin{equation}
Q_\eta \approx 54 y^2_i G^2_{\rm F} T^6 n_{\rm B} \approx 2.4 y^2_i
\times 10^{42}~{\rm erg}~{\rm cm}^{-3}~{\rm s}^{-1} \; ,
\end{equation}
where $\sigma = G^2_{\rm F} E^2$, $T = 30~{\rm MeV}$ and $n_{\rm B}
= \rho/m_N$ with $\rho = 3.0\times 10^{14}~{\rm g}~{\rm cm}^{-3}$
have been assumed. Note that all the neutrinos have been taken to be
relativistic and non-degenerate, which is an excellent approximation
for $\nu_\mu$ and $\nu_\tau$. For degenerate $\nu_e$, there will be
a blocking factor that suppresses the scattering rate, which has
been neglected in Eq. (4) for an order-of-magnitude estimate. The
energy loss should be small so as not to shorten the neutrino burst,
so we require the volume emission rate to be $Q_\eta \lesssim 3.0\times
10^{33}~{\rm erg}~{\rm cm}^{-3}~{\rm s}^{-1}$ and then obtain
\begin{equation}
y_i \lesssim 3.5\times 10^{-5} \; .
\end{equation}
As a matter of fact, there are additional contributions to the
production of $\eta$ bosons, such as $\nu_i + e^- \to \nu_i + e^- +
\eta$ and $\nu_i + \bar\nu_i \to \eta + \eta$. Hence the bound may
be slightly stronger if all the contributions are included.
One can then verify that the mean free path $\lambda_\eta \sim
10^8~{\rm cm}$ is much larger than the core radius, which is
consistent with the prerequisite that $\eta$ bosons can escape from
the core and carry away energies.

Based on the above discussions, one may take $y_i = 10^{-5}$ and
figure out the cross section of $\nu_i + \bar\nu_i \to \nu_j +
\bar\nu_j$ via $\eta$-exchange, and that via $Z^0$-exchange. It is
straightforward to get $\sigma_\eta \sim y^4_i/E^2$ for the former
case, while $\sigma_Z \sim G^2_{\rm F} E^2$ for the latter, where
$E$ is the neutrino energy. In the neutrinosphere with a typical
temperature $T = 10~{\rm MeV}$,  we further obtain
$\sigma_\eta/\sigma_Z \approx y^4_i/G_{\rm F}^2 E^4 \approx 10^{-4}$
for $y_i = 10^{-5}$ and $E = 3T = 30~{\rm MeV}$. Therefore, the
scalar-mediated neutrino interaction is too weak to equilibrate
supernova neutrinos of different species in the neutrinosphere and
thus cannot wash out the collective effects in supernova neutrino
oscillations \cite{Sher:2011mx}.

In conclusion, if the astrophysical bounds on the scalar-mediated
neutrino interaction are taken into account, the motivation for the
$\nu$2HDM becomes very weak in the sense that tiny Dirac neutrino
masses cannot be explained by a small vev but large Yukawa
couplings.

\vspace{0.2cm}

{\sl The author would like to thank Georg Raffelt for valuable
comments and suggestions, and Marc Sher for many interesting
discussions. This work was supported by the Alexander von Humboldt
Foundation.}


\end{document}